\documentclass[useAMS,usenatbib]{mn2e}
\usepackage{mathptmx}
\usepackage{graphicx}
\newcommand{\kms}{{\rm km~s^{-1}}}
\def\HI{{\sc H\,i}}
\def\subHI{H{\scriptscriptstyle I}}
\def\HII{{\sc H\,ii}}
\def\CO#1#2{$^{12}$CO ($J=#1{-}#2$)}
\title[H\,{\normalsize I} and CO observations of Arp~104]{H\,{\Large\bf I} and
  CO observations of Arp~104: a spiral--elliptical interacting pair}

\author[H. Cullen et al.]{H. Cullen,$^1$ P. Alexander,$^1$ D.~A. Green$^1$
  and K. Sheth$^2$\\
$^1$Astrophysics Group, Cavendish Laboratory, 19 J.~J.~Thomson Ave.,
    Cambridge CB3 0HE\\
$^2$Spitzer Science Center, California Institute of Technology,
   Pasadena, CA 91125, USA\\}

\date{2006 September 11}

\pagerange{\pageref{firstpage}--\pageref{lastpage}}
\pubyear{2006}
\begin{document}

\label{firstpage}

\maketitle

\begin{abstract}
We present data probing the spatial and kinematical distribution of both the
atomic ({\HI}) and molecular (CO) gas in NGC~5218, the late-type barred spiral
galaxy in the spiral--elliptical interacting pair, Arp~104. We consider these
data in conjunction with far-infrared and radio continuum data, and $N$-body
simulations, to study the galaxies interactions, and the star formation
properties of NGC~5218. We use these data to assess the importance of the bar
and tidal interaction on the evolution of NGC~5218, and the extent to which the
tidal interaction may have been important in triggering the bar. The molecular
gas distribution of NGC~5218 appears to have been strongly affected by the bar;
the distribution is centrally condensed with a very large surface density in
the central region. The $N$-body simulations indicate a timescale since
perigalacticon of $\sim 3 \times 10^{8}$ yr, which is consistent with the
interaction having triggered or enhanced the bar potential in NGC~5218, leading
to inflow and the large central molecular gas density observed. Whilst NGC~5218
appears to be undergoing active star formation, its star formation efficiency
is comparable to a `normal' SBb galaxy. We propose that this system may be on
the brink of a more active phase of star formation.
\end{abstract}

\begin{keywords}
 galaxies: ISM -- galaxies: interactions -- galaxies: individual: NGC~5218
 -- radio lines: galaxies
\end{keywords}

\section{Introduction}

Barred galaxies represent an important sub-sample of systems in any study of
galaxy interactions and induced star formation.  Bar formation is attributed to
one of two possible mechanisms; bars can form spontaneously as a result of an
intrinsic gravitational instability, or they can be produced by tidal
disturbance of the galactic potential during an interaction. Both mechanisms
are supported by numerical simulations. Massive stellar disks with small
velocity dispersions are intrinsically unstable to bar formation and will
develop a strong bar structure without external perturbation within a few
rotation periods \citep[e.g.,][]{1986MNRAS.221..213A}. Similarly, external
perturbations can induce bar structures in galactic disks that are stable in
isolation \citep*[e.g.,][hereafter MH96]{1987MNRAS.228..635N,
1990A&A...230...37G, 1996ApJ...464..641M}.

Observationally, \cite{1996ASPC...91..360K} used $K$-band images to study the
bar fraction in a sample of pairs, finding 80\% of the pairs to display signs
of bar formation. \cite*{1990ApJ...364..415E} also find a high fraction of
barred galaxies in binary systems ($50 \pm 5\%$), higher than in the field ($32
\pm 9\%$) or group environments ($28 \pm 2\%$). However,
\cite{2002AJ....124..782V}, using a sample of 930 galaxies found no evidence
for a dependence of bar frequency on galaxy environment, consistent with the
nature of the host galaxy being the most important factor in bar formation.
Taken together these results suggest that bar formation is strongly dependent
on the host galaxy, but that pair-wise interactions do induce bars in many
systems.

Galaxy bars represent a principal mode of disruption to the axisymmetric
potential of spiral galaxies, providing a mechanism of gas inflow toward the
central regions. The simulations of \cite{1994mtia.conf..372M} and MH96
indicate that bars can play a particularly important role in fly-by encounters,
where they provide a dominant trigger for central gas concentration and ensuing
star formation. In addition, the high proportion of galaxies in the local
universe hosting bars \citep{2000AJ....119..536E, 2004AAS...205.6102M} means
that the response of a barred potential to an external perturbation is also of
interest.  The relationship between bars and interactions and the evolution of
the ISM in barred systems is crucial in attaining a better understanding of the
star formation response of interacting galaxies.

Both theoretical (MH96) and observational \citep{1999ApJ...525..691S,
2005ApJ...632..217S} evidence exists pointing to enhanced molecular gas
concentration in barred systems (at least in early-type spirals). However, the
relationship between bars and star formation remains less clear. Whilst bars do
not appear to affect the disk-wide star-formation properties of galaxies, they
can be important in triggering enhanced star formation in the central regions
\citep[for a review see][]{1994mtia.conf..131K}. Star formation in barred
galaxies manifests itself in a diverse range of morphologies
\citep{1996ASPC...91...44P, 2000ApJ...532..221S, 2002AJ....124.2581S}. A number
of studies point to an enhanced level of star formation in the central region
of early-type barred systems, although similar enhancements are not observed in
late-type systems \citep{1996A&A...313...13H, 1997ApJ...487..591H}. This
distinction is consistent with molecular gas observations. Bars may well play a
particularly important role in the evolution of the most actively star-forming
systems. \cite{1996ASPC...91...44P} examined the $L_{\rm FIR}/L_{\rm B}$ ratio
in a sample of IR-bright galaxies and only detected a difference between the
barred and non-barred systems for the most actively star-forming galaxies.
Similarly, \cite{1999ApJ...516..660H} found a higher incidence of bars in
starburst galaxies, and \cite{1986MNRAS.221P..41H}, who looked at the
properties of intrinsically luminous galaxies, detected a mean IR luminosity
three times larger in barred systems than their unbarred counterparts.

In this paper we examine the spatial and kinematic distribution of both the
atomic and molecular gas associated with NGC~5218, the late-type barred spiral
galaxy in the spiral--elliptical interacting pair Arp~104. This is one of a
small sample of nine spiral--elliptical interacting pairs, which have been
studied by \cite{2003Ap&SS.284..503C, 2006MNRAS.366...49C} and
\cite{2005dmu..conf..353C}. Each pair has an {\HI} flux of at least
5~Jy~km~s$^{-1}$, and Arp 104 has a relatively large separation between the
galaxies, implying they are in the early state of interaction. We use these
data in conjunction with $N$-body simulations and data probing the star
formation properties of this system to assess the impact of the bar on
NGC~5218's evolution, and look at the extent to which the tidal interaction may
have been important in triggering its bar. In Section~\ref{observations} we
present details of our {\HI} and CO observations. In Section~\ref{results} we
discuss the star-formation properties and efficiency of NGC~5218 and present
the results of the simulations. In Section~\ref{bar_imp} we examine the
importance of the bar in this galaxy. In Section~\ref{discuss} we discuss our
results, looking first at the importance of the bar in terms of NGC~5218's
evolution and subsequently at the likelihood that the bar has been triggered by
the tidal interaction.

\section{Observations and data reduction}\label{observations}

\subsection{Optical morphology of Arp~104}

\begin{figure*}
\centerline{\includegraphics[clip=,
angle=180,width=0.75\textwidth]{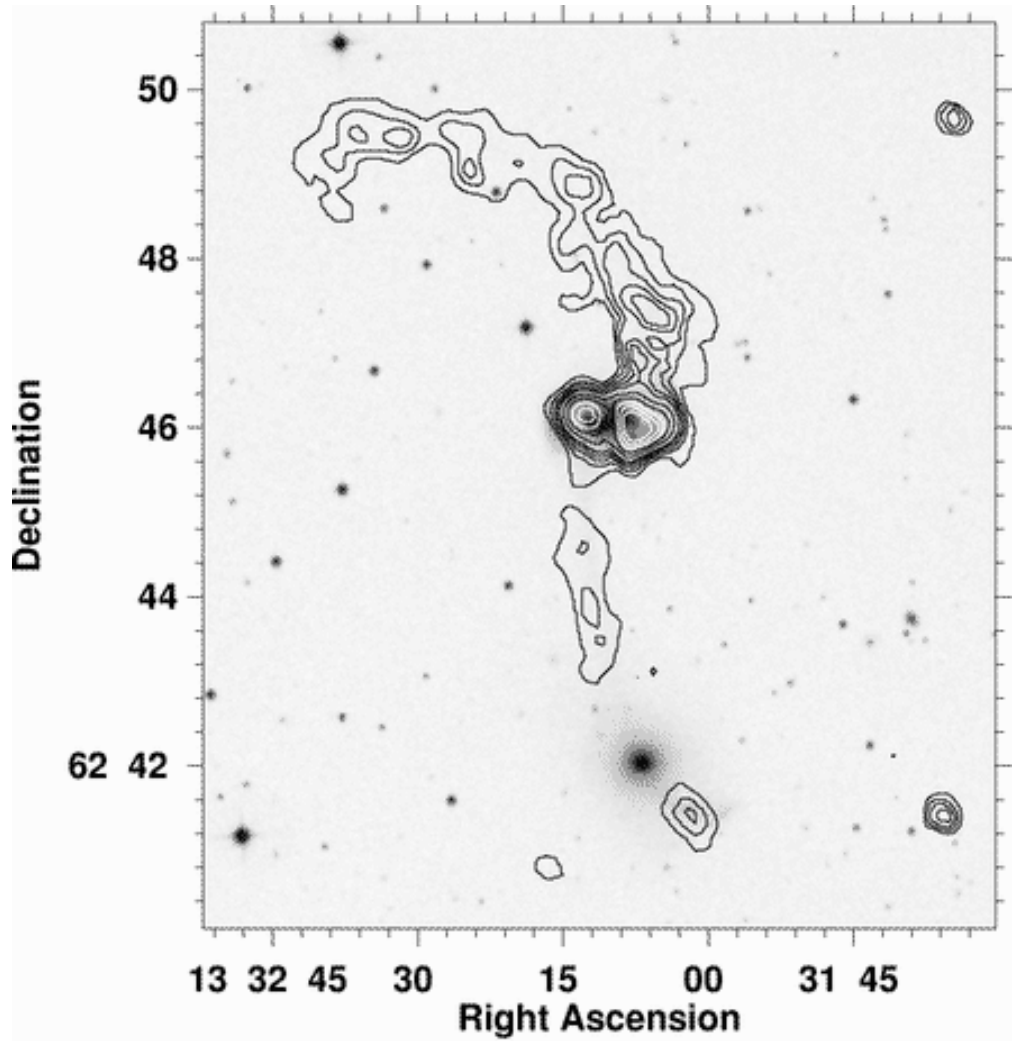}}
\caption{Integrated {\HI} column density from combined VLA C- and D-array data
($\theta_{\mathrm{FWHM}} = 27\farcs4 \times 24\farcs6$ at a position angle of
$53\fdg3$) for Arp~104 overlayed on a digitised sky survey $B$-band image.
Contour levels are (0.05, 0.075, 0.1, 0.125, 0.15, 0.2, 0.25, 0.3, 0.35, 0.4)
Jy~beam$^{-1}$ km~s$^{-1}$.}\label{a104hi_mom0.fig}
\end{figure*}

Arp~104, shown in Fig.~\ref{a104hi_mom0.fig}, consists of two interacting
galaxies, see NED\footnote{see {\tt http://nedwww.ipac.caltech.edu/}}: the
southern galaxy, NGC~5216, is an E0 pec ($2\farcm5 \times 1\farcm5$) and the
northern galaxy, NGC~5218, is an SB(s)b? pec ($1\farcm8 \times 1\farcm3$). At
an assumed distance of 41.1~Mpc (from a recessional velocity of 2880 $\kms$,
and using a Hubble constant of 70 $\kms$ Mpc$^{-1}$), the projected major axis
diameter ($D_{25}$) is 29.9~kpc ($2\farcm5$) for NGC~5216 and 21.5~kpc
($1\farcm8$) for NGC~5218. The projected separation of the two galaxies is
49.1~kpc. \cite*{1990AJ.....99..497S} have conducted a multicolour photometric
study of the optical tidal features of a sample of interacting galaxies,
including Arp~104.  They found the tidal features of Arp~104 are bluer than the
central regions of both NGC~5218 and NGC~5216, but similar to the colours of
the outer disk in NGC~5218.

\begin{figure*}
\centerline{\includegraphics[clip=,width=0.9\textwidth]{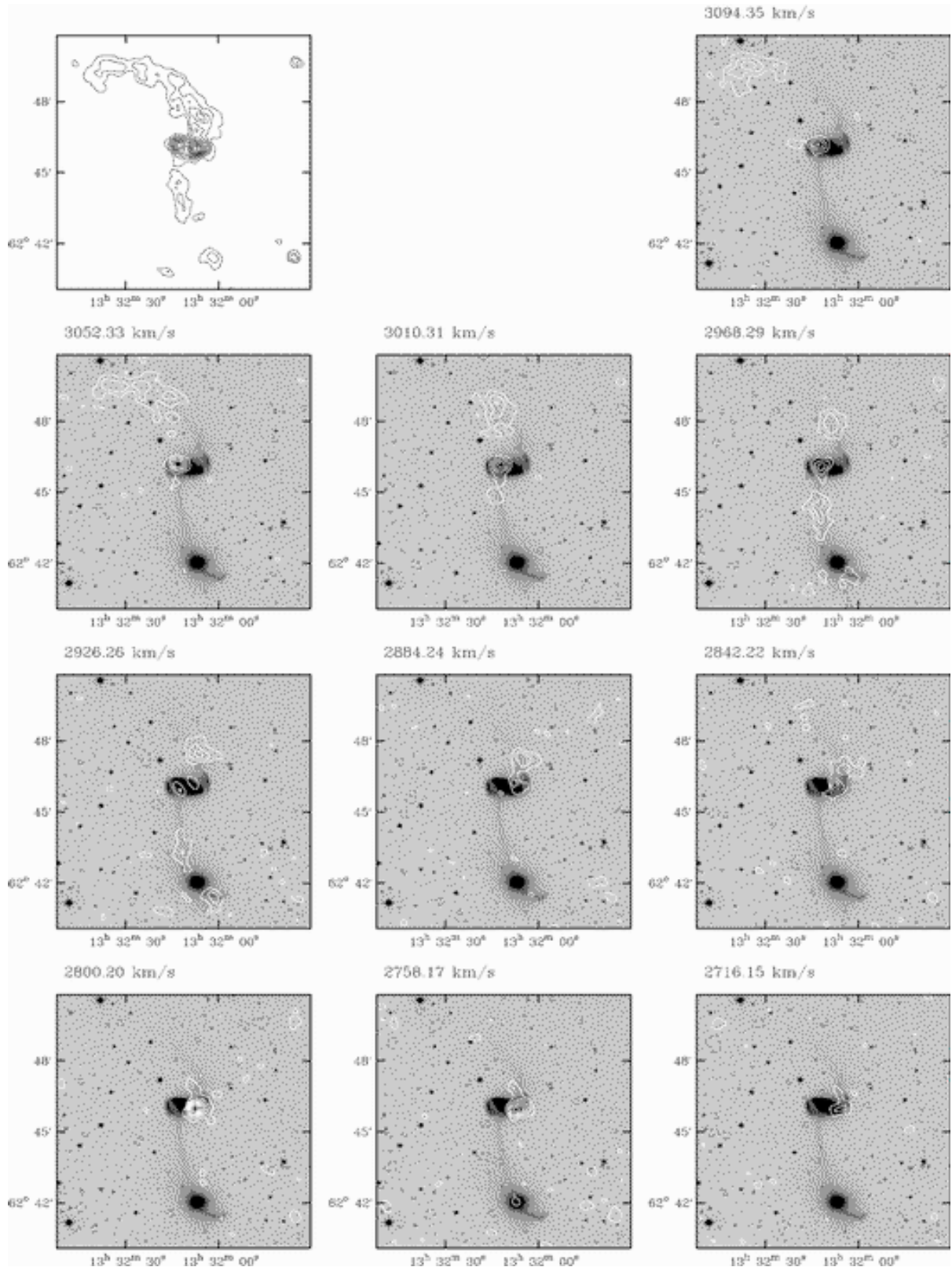}}
\caption{Channel maps for the combined VLA C and D-array data
($\theta_{\mathrm{FWHM}} = 27\farcs4 \times 24\farcs6$ at a position angle of
$53\fdg3$) for Arp~104 overlayed on a digitised sky survey $B$-band image. Maps
have been produced by averaging three adjacent channel maps together. Contour
levels are every 0.005 Jy~beam$^{-1}$ km~s$^{-1}$, omitting zero (negative
contour are dashed and black, positive contours are solid and
white).}\label{a104hi_channel.fig}
\end{figure*}

\begin{figure*}
\centerline{\includegraphics[clip=,angle=180,width=0.75\textwidth]{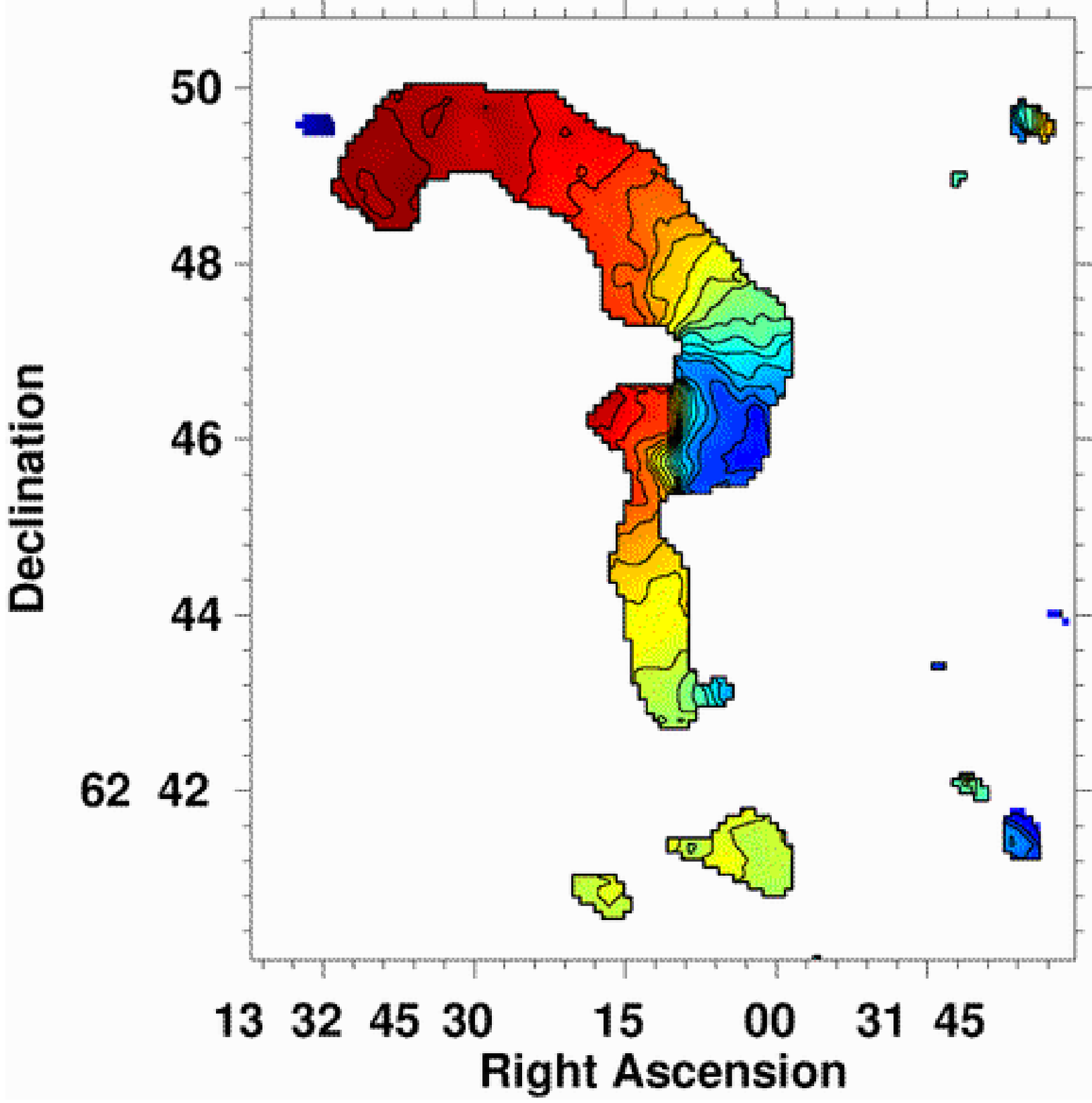}}
\caption{The intensity weighted mean velocity field of {\HI} emission from the
combined VLA C- and D-array data ($\theta_{\mathrm{FWHM}} = 27\farcs4 \times
24\farcs6$ at a position angle of $53\fdg3$) for Arp~104. The colour scale goes
from 2716 (blue) to 3093 $\kms$ (red) and contour levels are (2760, 2780, 2800
\ldots 3100) $\kms$.}\label{a104hi_mom1.fig}
\end{figure*}

\subsection{VLA 21-cm observations}

The {\HI} observations of Arp~104 were made with the VLA in C-array
configuration on 1994 December 24 and in D-array on the 1995 March 22. Both
datasets were retrieved from the VLA archive and analysed. For both the C- and
D-array observations, a correlator mode with online Hanning smoothing giving 63
channels with a velocity width per channel of 10.3 $\kms$ (48.8~kHz) was used.
Both sets of observations were centred on RA $13^{\rm h}$ $32^{\rm m}$
$09\fs36$, Dec $+62^\circ$ $44'$ $03\farcs3$ (J2000), at a heliocentric radial
velocity of 2900 $\kms$. The total on-source integration times were 258 and 236
minutes for the C and D-array observations respectively. In both cases
on-source scans were interleaved with those of the phase calibrator,
1400$+$621. For the C-array observations both 3C48 and 3C286 were observed as
flux calibrators during the observations, with assumed flux densities of 16.0
and 14.8 Jy respectively. For the D-array observations 3C286 was used as the
flux calibrator with an assumed flux density of 14.8 Jy.

Data reduction followed standard procedures using the National Radio Astronomy
Observatory (NRAO) Astronomical Image Processing System (AIPS). After flagging,
the data were binned to velocity resolution of 20.7 $\kms$ to improve the
signal to noise. Subtraction of the average of the line-free channels from all
channels was used to remove the continuum emission. AIPS task IMAGR was used to
produce naturally-weighted channel maps with an angular resolution $21\farcs3
\times 16\farcs4$ for the C-array data and $63\farcs9 \times 54\farcs4$ for the
D-array data.

\subsection{GMRT {\HI} absorption observations}

In addition to the {\HI} images from the VLA observations, NGC~5218 was also
observed with the Giant Metrewave Radio Telescope (GMRT) on 2003 October 16 to
obtain an {\HI} absorption spectra towards its nuclear continuum emission. The
GMRT is an aperture synthesis telescope comprising of thirty 45-m dishes
\citep{2002IAUS..199..439R}. The full width half maximum of the primary beam of
the telescope at 1.4~GHz is $\sim 25'$. The backend setup employed was a 30
station FX correlator with 128 spectral channels over a 4~MHz bandwidth (857
$\kms$), giving a resolution of 31.25 kHz. The observations were centred at a
heliocentric radial velocity of 2880 $\kms$. The total integration time on
source was 174 minutes. On-source scans were interleaved with those of the
phase calibrator, 1313$+$675. 3C286 was used as a flux calibrator with an
assumed flux density of 14.8 Jy. The pointing centre was the core of NGC~5218:
RA $13^{\rm h}$ $32^{\rm m}$ $10\fs50$, Dec $+62^\circ$ $46'$ $04\farcs0$
(J2000). These data were binned to a velocity resolution of 40 $\kms$ and
channel maps made using natural weighting, giving a spatial resolution of
$7\farcs72 \times 4\farcs60$. The flux density of the continuum source is
approximately 25~mJy, consistent with the FIRST observation
\citep{1997ApJ...475..479W}.

\subsection{\CO{1}{0} observations}

Observations of NGC~5218 in the \CO{1}{0} line at 115.27~GHz were made using
the 10-element Berkeley--Illinois--Maryland Association (BIMA) millimeter
interferometer. Two usable tracks were observed in C-configuration on 2003
April 28 and May 13. The BIMA correlator was configured to have a resolution of
1.56 MHz (4 $\kms$) over a total bandwidth of 368~MHz (959 $\kms$). The primary
beam at BIMA at this frequency is $100''$ FWHM. 3C84 and 1638$+$573 were
observed for passband and phase calibration respectively. Data calibration was
performed in Miriad \citep*{1995ASPC...77..433S}. Calibrated datasets were then
read into AIPS, combined, averaged to a spectral resolution of 20 $\kms$ and
images synthesised.

\subsection{\CO{2}{1} observations}

Observations of the \CO{2}{1} emission of NGC~5218 at 230.54~GHz were made with
the JCMT in `service' mode on 2001 November 30 and 2003 January 4. Observations
were made on a $9 \times 5$ grid in RA and Dec, with $10''$ spacing, centred at
RA $13^{\rm h}$ $32^{\rm m}$ $11\fs8$, Dec $+62^\circ$ $46'$ $05''$ (J2000).
The $10''$ spacing of grid points produced a fully-sampled map over the area
observed (JCMT beam $20\farcs8$ at 230.54~GHz). Sky subtraction was achieved by
beam switching to a point $80''$ in declination from the reference position.
Observations were centred on a heliocentric radial velocity of 2880 $\kms$ with
a bandwidth of 920~MHz (1200 $\kms$), split into channels separated by 625 kHz.
The spectra were binned to a velocity resolution of 20.9 $\kms$ to match that
of the VLA data.

Standard calibration observations were made at the same time as the
observations of NGC~5218. Data calibration followed the standard JCMT
procedure, adopting the beam efficiency and forward scattering efficiency on
the JCMT website yielding an $\eta_{\rm fss}$ of 0.77. To convert between main
beam temperature and flux it was assumed that the \CO{2}{1} emission could be
modelled as a point source. Whilst this is not strictly the case, it allows a
lower limit to be placed on the \CO{2}{1} flux for NGC~5218. The conversion
between flux and main beam temperature, assuming a point source, is given by
$(S/{\rm Jy}) = 18.4 (T_{\rm mb}/{\rm K})$ (assuming the FWHM of the JCMT beam
at 230~GHz is $20\farcs8$).

\section{Results}\label{results}

\subsection{Star-formation rate}\label{5218_starformrate.sec}

The flux densities of NGC~5218 at 12, 25, 60 and 100 $\mu$m, taken from the
IRAS High Resolution Image Restoration Atlas \citep{1997ApJS..111..387C}, are
0.36, 0.92, 7.14 and 14.38 Jy respectively. This yields a far-infrared (FIR)
luminosity \citep[$40{-}122$ $\mu$m,][]{1988ApJS...68..151H} of
$8.37\times10^{36}$ W ($2.2 \times 10^{10}$ L$_{\odot}$). Using the calibration
of \cite{2003ApJ...586..794B} we obtain a FIR star-formation rate (SFR) of
5.3~M$_{\odot}$ yr$^{-1}$. The infrared luminosity ($8{-}1000$ $\mu$m) for
NGC~5218 is $3.96 \times 10^{10}$ L$_{\odot}$ \citep{1987PhDT.......114P}.

Both NGC~5218 and NGC~5216 have detected radio continuum emission recorded by
the NVSS survey \citep{1998AJ....115.1693C}, with 30.4 and 31.8 mJy
respectively. The NVSS flux density of NGC~5218 corresponds to a luminosity of
6.16$\times10^{21}$ W Hz$^{-1}$.  Using the radio SFR calibration of
\cite{2003ApJ...586..794B} this yields a SFR of 3.4 M$_{\odot}$ yr$^{-1}$, a
factor of $\sim 2$ below the SFR obtained using the calibration of
\cite{1992ARA&A..30..575C} (7.6 M$_{\odot}$ yr$^{-1}$). For the purposes of
this work we adopt the FIR estimated SFR of 5.3 M$_{\odot}$ yr$^{-1}$.

\subsection{Atomic Hydrogen}\label{a104_hi.sec}

The combined VLA C- and D-array {\HI} map of Arp~104
(Fig.~\ref{a104hi_mom0.fig}) reveals a tidal tail to the north-east of the
late-type system (NGC~5218), and a tidal bridge extending southward toward the
early-type companion (NGC~5216). Figure~1, like the other {\HI} and CO results
shown here, has not been corrected for the primary beam of the telescope, which
is much larger than the scale of the emission that is seen. The tidal tail to
the north-east forms a quarter-circle arc with a radius of approximately $4'$,
corresponding to a physical radius of 48~kpc. The tidal bridge to the south is
approximately $5'$ long, corresponding to a physical length of 60~kpc. The
bridge is oriented directly toward the elliptical, NGC~5216, with a small tail
extending beyond the elliptical to the south-west.

Using the lower resolution D-array data we detect an {\HI} flux above a level
of 0.035 Jy beam$^{-1}$ $\kms$ over an area of approximately 37 arcmin$^{2}$.
At a distance of 41~Mpc, this corresponds to a physical area of 5360~kpc$^{2}$.
The detected flux in this low resolution map peaks at 0.98 Jy beam$^{-1}$
$\kms$, falling to around 0.05 Jy beam$^{-1}$ $\kms$ in the more extended
regions of the tidal tail.

The total {\HI} flux of the Arp~104 system, obtained by adding all the flux in
the low-resolution channel maps is $7.0 \pm 0.5$ Jy $\kms$, which is in good
agreement with the single-dish flux, $7.7 \pm 2.4$ Jy $\kms$
\citep{1998A&AS..130..333T}. The {\HI} mass, given by
\[
  M_{\subHI} = 2.36 \times 10^{5} D^{2} \int S \, {\rm d}v
     \quad {\rm M_{\odot}},
  \label{HI_mass}
\]
is $(2.8 \pm 0.2) \times 10^{9}$ M$_{\odot}$.

Approximately 30\% of the {\HI} emission detected in the D-array map appears to
be associated with the late-type system, NGC~5218, with the remainder forming
an extended tidal tail. This is consistent with the trend found by
\cite{1996AJ....111..655H}. For a sample of four galaxies (Arp 295b, NGC 4676a,
NGC 4676b and NGC 520) considered to lie in the central region of the merger
sequence, they find $\sim 35\%$ of the total atomic gas emission comes from the
inner regions of the system; this percentage falls dramatically for more
advanced mergers.

Channel maps of the {\HI} emission are shown in Fig.~\ref{a104hi_channel.fig},
whilst Fig.~\ref{a104hi_mom1.fig} displays the intensity weighted mean velocity
field of Arp~104. {\HI} emission is detected in channels 7--25 corresponding to
the velocity range $2716{-}3094$ $\kms$. The northern tidal tail has a velocity
dispersion of $\sim 270$ $\kms$ with velocities in the range $2821{-}3094$
$\kms$. The tidal bridge to the south has a slightly smaller velocity
dispersion of $\sim 210$ $\kms$, with velocities in the range $2800{-}3010$
$\kms$. The largest velocity dispersion is observed for the gas still
associated with NGC~5218. Here velocities range from $2716{-}3094$ $\kms$ with
the western side blueshifted.

\begin{figure}
\centerline{\includegraphics[clip=,angle=270,width=\columnwidth]{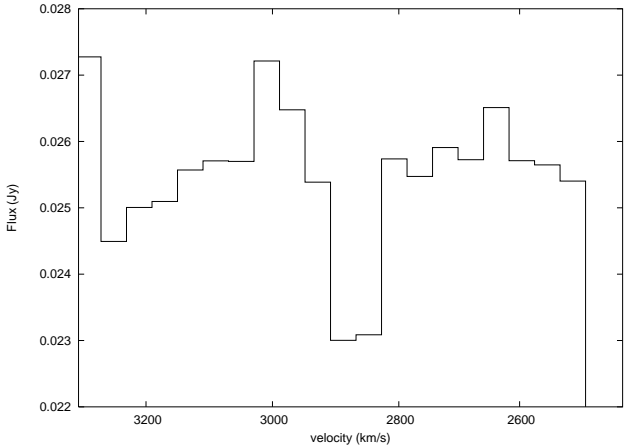}}
\caption{{\HI} absorption spectra towards the central continuum source of
NGC~5218. Data at a spatial resolution of $7\farcs72 \times 4\farcs60$ at a
position angle of $-40\fdg1$ binned to a velocity resolution of 40
$\kms$.}\label{5218_hiabss.fig}
\end{figure}

The central region of NGC~5218 is seen in absorption due to the presence of a
bright radio continuum source in NGC~5218 (see Fig.~\ref{5218_hiabss.fig}). The
absorption profile is centred on a velocity of 2884 $\kms$. The detected {\HI}
emission believed to be still bound to the late-type system is distributed
approximately symmetrically about this velocity. In contrast, emission from
tidal material is mostly redshifted.

The depth of the {\HI} absorption feature give an optical depth of $0.10 \pm
0.02$ with a linewidth of 80 $\kms$.  Assuming a spin temperature of 100~K,
this corresponds to an {\HI} column density of $(1.5 \pm 0.3) \times 10^{21}$
cm$^{-2}$. If the continuum source is located in the centre of the galaxy we
would observe approximately half of the {\HI} along the line of sight in
absorption. This implies a total {\HI} column density through the centre of the
disk of $(3.0 \pm 0.6) \times10^{21}$ cm$^{-2}$.

\subsection{Molecular Gas}\label{5218_co.sec}

\begin{figure}
\centerline{\includegraphics[clip=, angle=90,
width=\columnwidth]{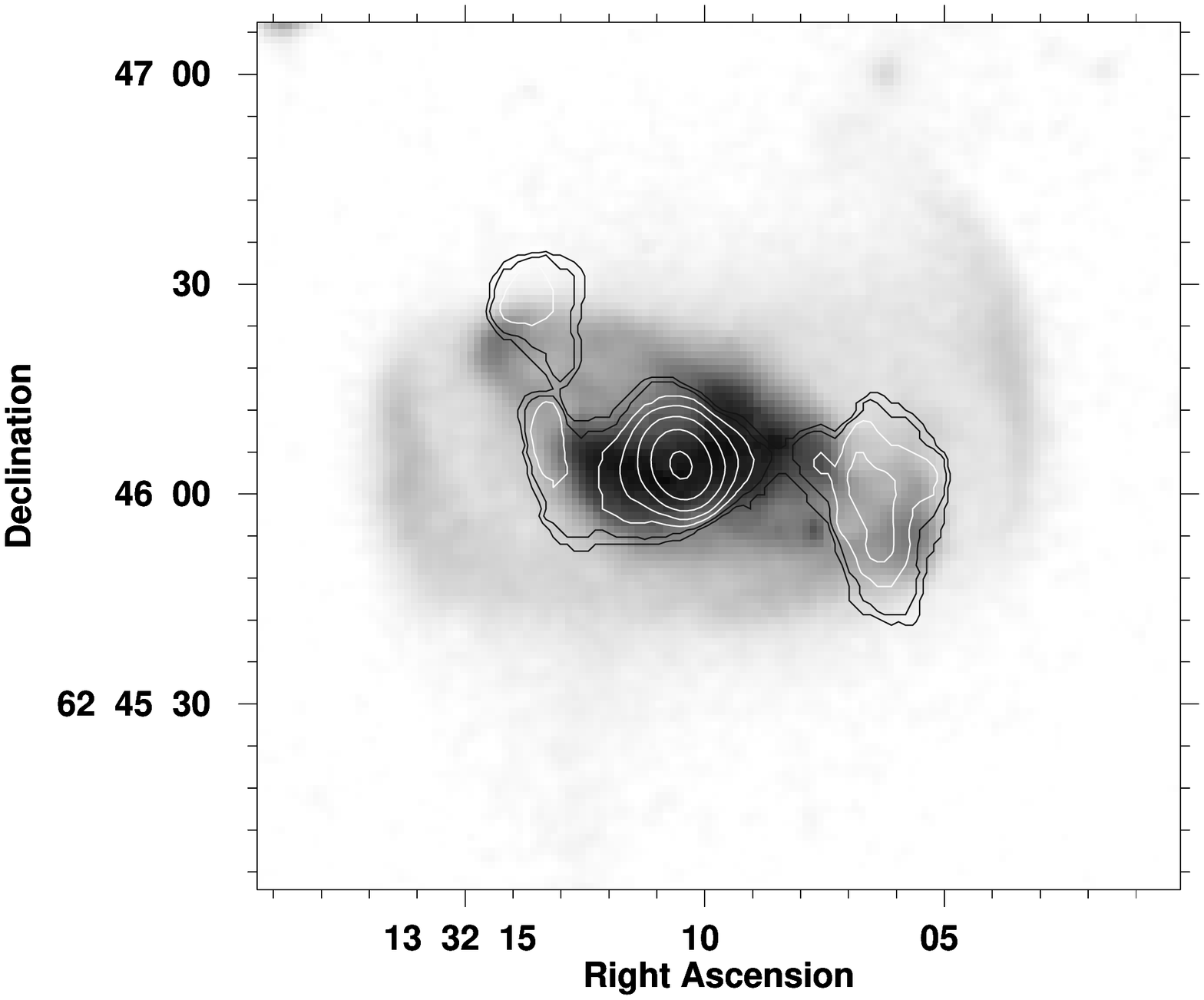}}
\caption{Integrated CO emission from BIMA \CO{1}{0} ($\theta_{\mathrm{FWHM}} =
8\farcs81 \times 5\farcs30$ at a position angle of $13\fdg0$) observations of
NGC~5218 overlayed on a digitised sky survey $B$-band image. The contour levels
are (2, 4, 8, 16, 32, 64, 128) Jy~beam$^{-1}$ $\kms$.}\label{5218co10_mom0.fig}
\end{figure}

Naturally weighted BIMA data were used to produce an image cube with a
synthesised beam of $\theta_{\mathrm{FWHM}} = 8\farcs8 \times 5\farcs3$,
corresponding to a physical scale of $\sim $1.4~kpc.
Figure~\ref{5218co10_mom0.fig} shows the integrated \CO{1}{0} map overlayed on
digitised sky survey $B$-band image. The CO data reveal four distinct peaks.
The central peak is significantly brighter than the peaks observed offset to
the east and west, its emission exceeding that in the eastern and western peaks
by a ratio of $10:1$ and $7:1$ respectively.

Whilst the molecular gas emission in NGC~5218 is clearly centrally condensed
there is also a sizeable extended component. The central peak itself is
extended in the easterly direction, well aligned with the optical bar axis.
Examining the subsidiary peaks in the molecular gas, there is a clear peak
located toward the eastern end of the bar, whilst at the western end there is a
larger area of emission extending from the bar end toward a region of star
formation in the spiral arm. There is similarly a peak in the eastern region of
the galaxy associated with a region of star formation in the eastern spiral
arm.

The integrated \CO{1}{0} flux for NGC~5218 was obtained from the primary beam
corrected naturally weighted image cube. The molecular gas mass was estimated
in solar masses from the integrated \CO{1}{0} flux using the empirical
relation:
\[
  M_{\rm H_{2}} = 1.6 \times 10^{9} \lambda^{2} D^{2} \int S \, {\rm d}v
   \quad {\rm M_{\odot}}\label{H2_mass.ch5}
\]
where $\lambda$ is the wavelength in metres, $D$ is the distance in megaparsecs
to the source, $\int S \, {\rm d}v$ is the integrated CO flux in Jy km~s$^{-1}$
and CO-to-H$_{2}$ conversion factor, $X \equiv N({\rm H_2})/I_{\rm CO} = 2.8
\times10^{20}$ cm$^{-2}$ K~km~s$^{-1}$ has been assumed
\citep{1986A&A...154...25B}. The observed integrated flux of $375 \pm 75$ Jy
$\kms$ is in good agreement with single-dish observations of
\cite{1999AJ....118..145Z}, who obtained $363 \pm 27$ Jy $\kms$, which implies
that the BIMA observations are not missing any flux. Our flux gives a total
H$_{2}$ mass of $(6.9 \pm 1.4) \times 10^{9}$ M$_{\odot}$.

The gas surface density in NGC~5218 is given by \citep{2003PASJ...55...87S}:
\[
  \left(\frac{\Sigma_{\rm H_{2}}}{\rm M_{\odot} \rm pc^{-2}}\right) =
  5.0 \times 10^{2} \cos(i)
  \left( \frac{I_{\rm CO}}{\rm Jy\ \kms\ arcsec^{-2}} \right),
\]
and $\Sigma_{\rm gas} = 1.36\Sigma_{\rm H_{2}}$, where $i$ is the galaxy
inclination and the factor 1.36 accounts for elements other than hydrogen
\citep[e.g.,][]{1973asqu.book.....A}. To estimate the peak central gas surface
density from our \CO{1}{0} data we re-made the image cubes applying uniform
weighting, giving a spatial resolution of $5\farcs48 \times 4\farcs58$. These
data gave a peak central gas surface density of $\sim 2000$ M$_{\odot}$
pc$^{-2}$. Additional observations of the molecular gas in this galaxy have
been made by \cite*{EvertOlsson_ngc5218} at higher resolution ($3\farcs75
\times 3\farcs54$) using the OVRO millimetre array. Using these data they
observe a peak gas surface density of $\sim 3000$ M$_{\odot}$ pc$^{-2}$.

\begin{figure}
\centerline{\includegraphics[clip=,angle=90,width=\columnwidth]{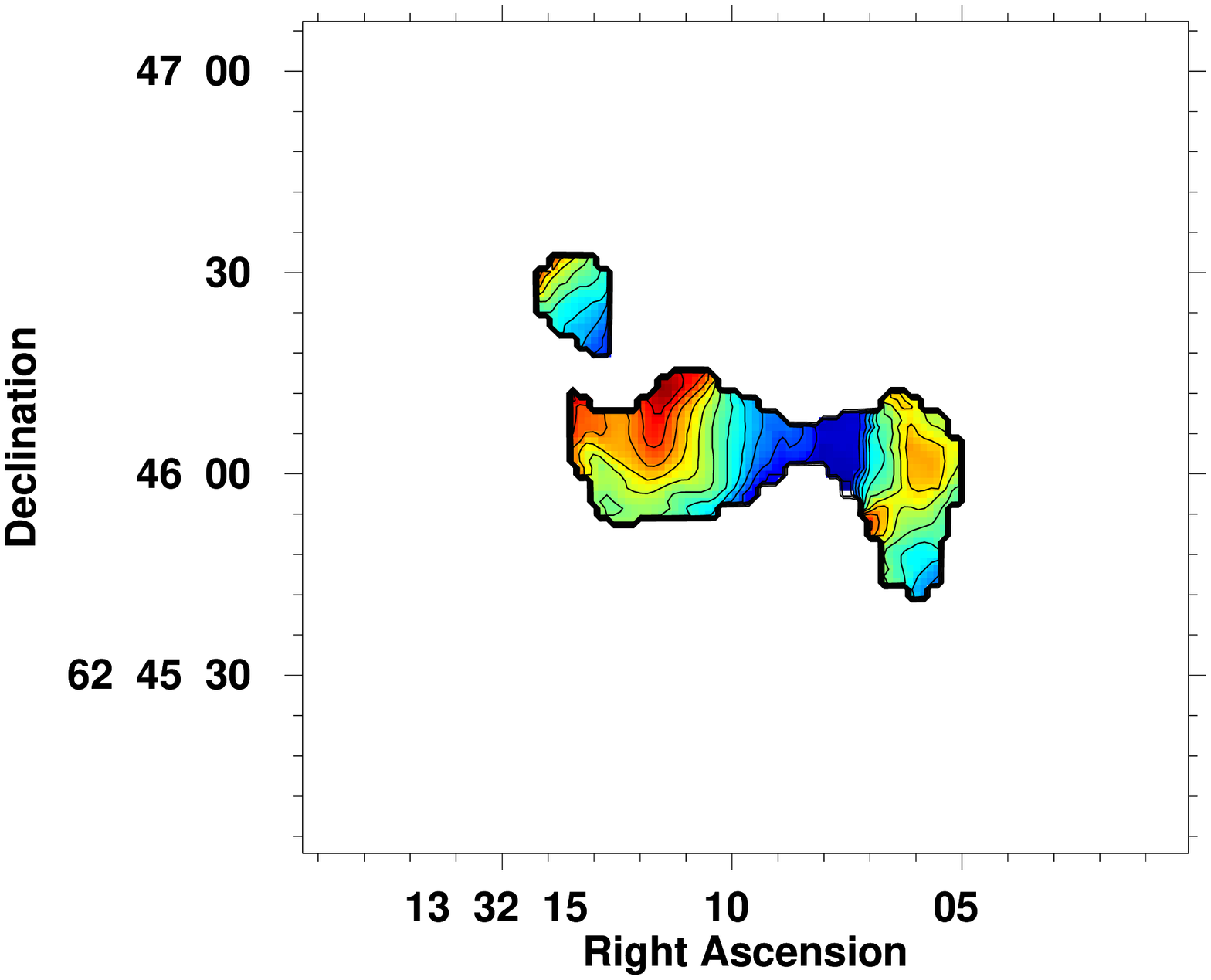}}
\caption{{\bf Left:} The intensity weighted mean velocity field of the BIMA
\CO{1}{0} data ($\theta_{\mathrm{FWHM}} = 8\farcs81 \times 5\farcs30$ at a
position angle of $13\fdg0$) for NGC~5218 The colour scale goes from 2716
(blue) to 3016 $\kms$ (red) and contour levels are (2730, 2750, 2770 \ldots
3050) $\kms$.}\label{5218co10_mom1.fig}\label{5218co10_vel.fig}
\end{figure}

\begin{figure}
\centerline{\includegraphics[clip=,angle=90,width=\columnwidth]{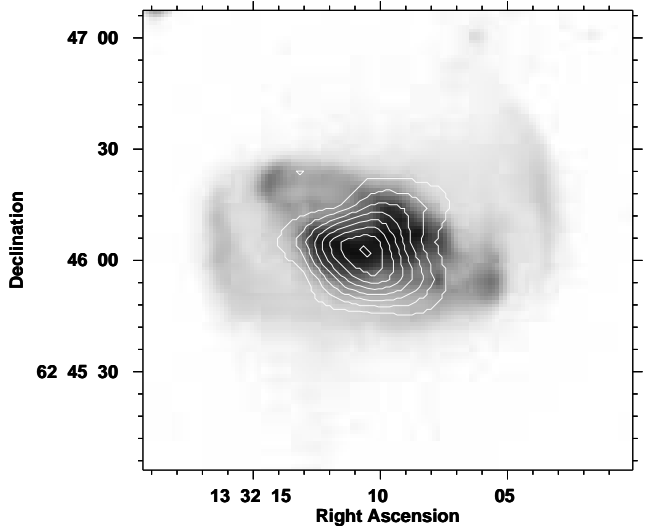}}
\caption{Integrated \CO{2}{1} emission ($\theta_{\mathrm{FWHM}} = 20\farcs8$)
from JCMT observations of NGC~5218 overlayed on a digitised sky survey $B$-band
image.  The contour levels are (2, 6, 10, 14, 18, 22, 26, 30) K
$\kms$.}\label{5218co21_mom0.fig}
\end{figure}

The intensity weighted mean velocity field of the BIMA \CO{1}{0} data is shown
in Fig.~\ref{5218co10_vel.fig}. These data are used later to fit a rotation
curve for NGC~5218. The \CO{2}{1} emission observed with the JCMT is shown in
Fig.~\ref{5218co21_mom0.fig}. CO emission is detected over a region
approximately $50'' \times 30''$.  For comparison, we convolved the \CO{1}{0}
data down to the same resolution as the JCMT data ($20\farcs8$) and re-gridded
the datasets to a common grid; the two datasets appeared consistent over their
overlap region.

We measure the \CO{2}{1}/\CO{1}{0} line ratio at the centre of NGC~5218 (RA
$13^{\rm h}$ $32^{\rm m}$ $10\fs2$, Dec $+62^\circ$ $46'$ $4\farcs85$) to be
$0.8 \pm 0.2$. This line ratio is similar to that found in Orion
\citep{1994ApJ...425..641S} and is consistent with the presence of giant
molecular clouds that have a kinetic temperature, $T_{\rm k} \sim 20$~K and
density, $n({\rm H}_{2}) \geq 10^{3}$ cm$^{-3}$. This ratio is similar to that
observed by \cite{1992A&A...264..433B} in the centre of normal spiral galaxies
($0.89 \pm 0.06$). The indication of a normal molecular gas environment at the
centre of NGC~5218 is corroborated by the $^{13}$CO ($J=1{-}0$)/\CO{1}{0} line
ratio of $9.0 \pm 1.3$ \citep{1991A&A...249..323A}, which is comparable to the
value observed in our Galaxy when averaged over the inner 300~pc.

\subsection{The relative gas content of NGC~5218}\label{gas_content.sec}

The \CO{1}{0} data for NGC~5218 give a molecular gas mass of $(6.9 \pm
1.4)\times10^{9}$ M$_{\odot}$ (see Section~\ref{5218_co.sec}). The mass of
atomic gas still bound to the late-type system is $(8.4 \pm 0.8)\times10^{8}$
M$_{\odot}$, while the total atomic gas mass associated with the Arp~104 system
is $(2.8 \pm 0.2)\times10^{9}$ M$_{\odot}$ (see Section~\ref{a104_hi.sec}).

Allowing for differences in the H$_{2}$-to-CO conversion factor, we compare the
molecular-to-atomic gas ratio of NGC 5218 to galaxies in the literature.
Including all the atomic gas associated with the extended tidal features in NGC
5218, we obtain a molecular-to-atomic ratio, in terms of $\log (M_{{\rm
H}_2}/M_{\rm HI})$ of 0.40, which is significantly larger than that observed in
both normal ($-0.45 \pm 0.07$, from \citep{1990A&A...233..357C}, and $-0.15$
from \citep{2003A&A...405....5B}) an peculiar SBb systems ($-0.18 \pm 0.15$,
\citep*{2004A&A...422..941C}). All mean ratios quoted in the literature
indicate SBb systems have less molecular gas than atomic gas; in contrast,
NGC~5218 appears to have at least twice as much molecular than atomic gas.

\subsection{The dynamics of the interaction}\label{104_sim.sec}

The spatial extent of Arp~104's tidal tails can be used to make an approximate
estimate the timescale of the interaction. The semi-circular tail extending to
the north of NGC~5218 is $\sim 6\farcm3$, which corresponds to a physical
distance of $\sim 75$~kpc, whereas the tidal bridge to the south is $\sim 5'$,
or approximately 60~kpc in extent. The two tidal extensions have comparable
lengths, indicative of similar formation timescales. If we assume that the
atomic gas associated with NGC~5218 was not accelerated during the encounter,
then the velocities of the gas will be comparable to those prior to the
interaction.  The velocity of the atomic gas that has travelled approximately
70~kpc to the tip of the tidal tails can then be estimated from the rotational
velocity of the system and the orbital velocity of the galaxy, $v_{\rm gas} =
(\Delta V)/2 + v$, where $v$ is the rotational velocity of the galaxy from
which the tidal tail originated and $\Delta V$ is the relative radial velocity
of the two interacting galaxies. The interaction between NGC~5218 and NGC~5216
appears to be largely in the plane of the sky; the galaxies have very similar
systemic velocities so we assume a relative radial velocity of zero. We
estimate the rotational velocity of NGC~5218 to be $\sim 247$ $\kms$ (corrected
for inclination). The spatial extent of the tidal tail and the estimated
velocity of the gas imply a time since perigalacticon of approximately $2.8
\times 10^{8}$ yr.

As a further probe of the dynamical evolution of the Arp~104 interaction we
have modelled the observed {\HI} distribution using collisionless $N$-body
simulations which provide a better constraint on the dynamics. Our
observational data place some immediate constraints on the simulations,
including: the absolute masses of the galaxies involved and their mass ratio;
the relative velocity difference of the line of sight velocities; the relative
orientation of the galaxies; the physical separation of the galaxies and the
length and orientation of the tidal tails.

Dynamical, $B$-band and $K$-band mass estimates for NGC~5218 and 5216, where
appropriate, are given in Table~\ref{104_mass.tab}. For the dynamical mass
estimate of NGC~5218 a spherical mass distribution and inclination angle of
$49\fdg5$ (taken from the HyperLeda database\footnote{see: {\tt
http://leda.univ-lyon1.fr/}}) were assumed. The $B$-band mass estimates used a
solar absolute $B$-band magnitude of 5.46 and assumed a mass-to-light ratio of
4 \citep{1994ARA&A..32..115R} for the late-type system and between 6
\citep{1991MNRAS.253..710V} and 10 \citep{1979ARA&A..17..135F} for early-type
system. The $K$-band mass estimates assume a stellar $K$-band absolute
magnitude of 3.32 and adopt a mass-to-light ratio of $1:1$ for both the early
and late-type systems \citep{2003ApJS..149..289B}. Both the $B$-band and
$K$-band estimates are from the total emission magnitudes of the galaxies.
Whilst these mass estimates are subject to a large degree of uncertainty, they
nonetheless provide a useful starting point for simulations of this system.

Using these constraints a series of $N$-body simulations were run using an
in-house tree code \citep{1999MNRAS.308..364C}.  Preliminary, restricted
$N$-body simulations were used to explore the parameter space. Simulations
started with zero energy (parabolic) orbits.  The restricted $N$-body
simulations allowed exploration of a range of possible perigalactic
separations. Having established a parameter range within which simulations
produced tidal features that were a good representation of those observed,
further, self-consistent simulations were run to confirm the validity of the
selected parameters. The criteria for the `best-fit' model was a simulated mass
distribution in which the projected separation of the galaxy pair was a good
match to that observed and the morphology and velocity profile of the simulated
galaxies most closely matched that of the {\HI} distribution.

The self-consistent $N$-body simulations of galactic systems used two types of
collisionless particles to simulate the stellar and dark matter; we ignore
hydrodynamical effects in the formation of the tidal tails. Initial conditions
for the simulated late-type galaxies are determined by a technique developed by
\cite{1993ApJS...86..389H}. Each late-type galaxy consists of a dark matter
halo, a central stellar bulge, and a rotating disk. The particles are then
positioned, following a spatial density profile modelled on observed profiles.
The halo, disk and bulge components were assigned the mass ratio $1.0:1.0:0.3$,
the halo and bulge components we each composed of 16384 particles, whilst the
disk component contained 32768 particles. The early-type system was simulated
using the NEMO code \citep{1995ASPC...77..398T} to construct a Plummer model.
In each case the model galaxies were evolved under gravity to confirm their
stability prior to use in simulations.

\begin{table}
\caption{Galaxy Mass Estimates}\label{104_mass.tab}
\begin{tabular}{@{}cccc}
\hline
Galaxy   &         $B$-band         &      $K$-band     &      Dynamical       \\
         &      Mass Estimate       &   Mass Estimate   &     Mass Estimate    \\
         &      (M$_{\odot}$)       &   (M$_{\odot}$)   &     (M$_{\odot}$)    \\ \hline
NGC~5218 &       $8 \times 10^{10}$ & $7 \times10^{10}$ & $1.3 \times 10^{11}$ \\
NGC~5216 & $(4{-}7) \times 10^{10}$ & $4 \times10^{10}$ & --                   \\ \hline
\end{tabular}
\end{table}

\begin{figure*}
\centerline{\includegraphics[clip=,angle=0,width=12cm]{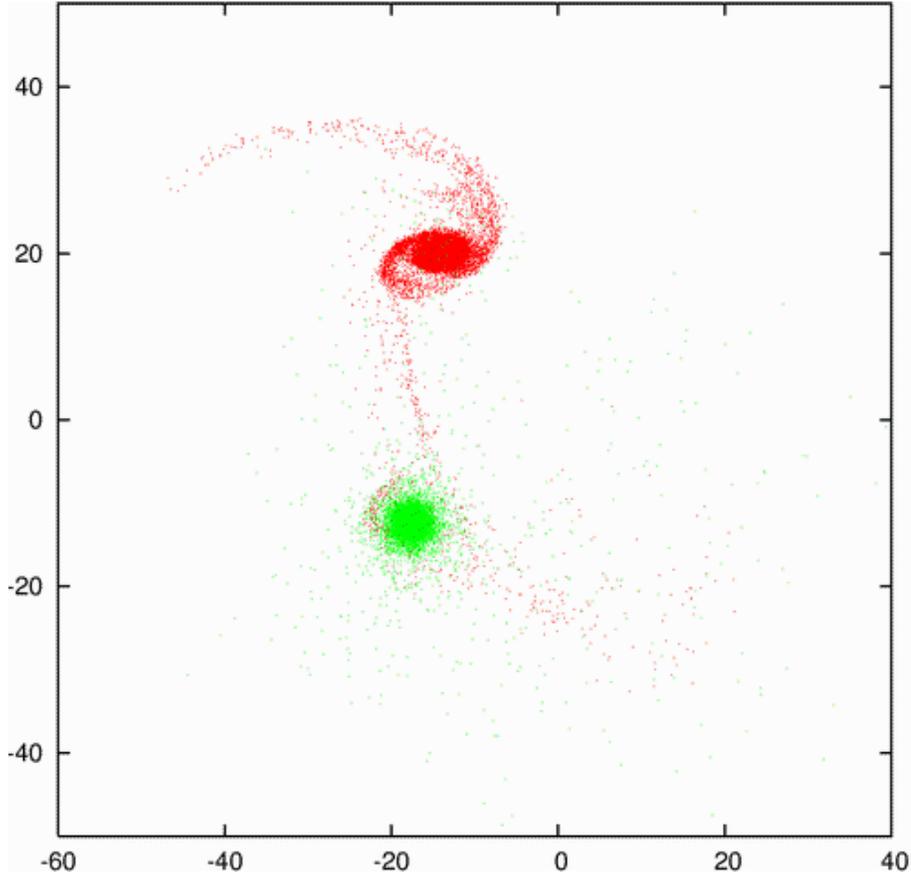}}
\caption{The distribution of collisionless particles obtained from a simulation
of the encounter between NGC 5216 and NGC 5218. Red particles indicate the disk
particles of the late-type component and green particles indicate the particles
comprising the early-type component. The orbital plane of the galaxies is in
the plane of the figure. The axes indicate distance in
kpc.}\label{sim_7_4_t1100.fig}
\end{figure*}

A series of simulations were run using models for an elliptical and spiral
galaxy on zero energy (parabolic) or small positive energy (hyperbolic) orbits.
A number of mass scalings were explored. These included, adoption of the
$K$-band mass estimates, adoption of the $K$-band mass ratio but using the mass
scale defined by the dynamical mass estimate for NGC~5218 and adoption of the
dynamical mass estimate for NGC~5218 and the $K$-band mass estimate for
NGC~5216. The last of these combinations implied a mass ratio of the late to
early-type galaxies of $4:1$. These simulations produced short tidal tails,
unlike those observed in the {\HI} data. On the basis of this work we suggest
that the mass ratio of the two galaxies is closer to $2:1$, as implied by the
$K$-band mass ratio.

All the simulations run using a $2:1$ mass ratio for the late to early-type
galaxies produced a tidal-tail morphology similar to that observed in the {\HI}
data. The results obtained appeared relatively robust. Several simulations were
run, with an initial disk inclination of 0, 30, 60 and 90 degrees. The best-fit
set of parameters were based on a parabolic interaction with a perigalactic
separation of 12.5~kpc, using galaxies with a masses approximated using the
$K$-band mass-to-light ratio; the simulated data, for an initial disk
inclination of 60 degrees, are shown in Fig.~\ref{sim_7_4_t1100.fig}. The time
from perigalacticon to the point at which the simulated data best represented
the {\HI} observations was $\sim 50$ time-steps, corresponding to a timescale
of $\sim 3.3 \times 10^{8}$ yr. This is in good agreement with the crude
dynamical estimate derived above.

\section{Analysis of the bar in NGC~5218}\label{bar_imp}

\subsection{Rotation Curve}

The first step in the analysis of the bar is to attain a rotation curve for
NGC~5218. The rotation curve was established predominately using the VLA {\HI}
data. Using the channel maps (Fig.~\ref{a104hi_channel.fig}), spatial and
velocity selection were use to isolate gas which is dynamically associated with
NGC~5218, this corresponded to gas out to a radius of approximately 9~kpc.
NGC~5218 contains a bright central radio continuum source and as a result {\HI}
in the nuclear region is seen in absorption, making it difficult to fit a
rotation curve to the {\HI} data in this region. However, in the central region
our \CO{1}{0} observations provide a reliable estimate of the rotation curve.
We therefore use {\HI} data for radii beyond about 2.5~kpc and CO data within
this radius. The \CO{1}{0} observations have the additional advantage that they
are higher resolution than the {\HI} data, reducing the problems of
beam-smearing in the centre of the galaxy \citep{1999ApJ...523..136S}.

Rotation curves for NGC~5218 were derived from the intensity-weighted velocity
field of the {\HI} and CO data using the AIPS task GAL. For both the {\HI} and
CO datasets we used the highest resolution images, so as to minimise the effect
of beam smearing. For the {\HI} data this corresponded to a spatial resolution
$13\farcs3 \times 10\farcs7$, and for the CO data the resolution was $8\farcs8
\times 5\farcs3$. The rotation curves were derived using a tilted-ring model
\citep{1973MNRAS.163..163W}. This method approximates the galaxy as a set of
concentric rings of gas at increasing radii, characterised by a circular
velocity, position angle and inclination angle. Attempts to derive the rotation
curve leaving the dynamic centre and inclination angle as free parameters
yielded unphysical results. The dynamical centre was therefore calculated by
performing a global rotation curve fit to {\HI} datasets at three different
resolutions for both Brandt and Exponential rotation curves independently.  The
dynamical centre was found to be stable across all fits and was therefore fixed
for the purposes of the tilted-ring curve fitting. Optical data were used to
fix the inclination angle which was taken from the Hyperleda
database\footnote{See {\tt http://leda.univ-lyon1.fr/}}.

\begin{figure}
\centerline{\includegraphics[clip=,angle=270,width=\columnwidth]{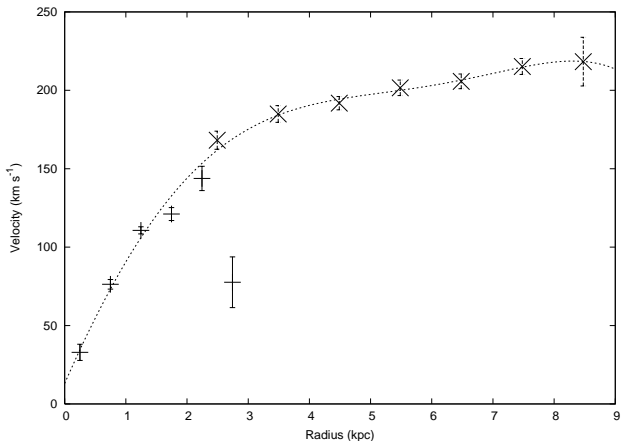}}
\caption{The model circular velocity de-projected on the galaxy plane as
obtained by fitting a tilted-ring model to the combined molecular (plus signs)
and atomic (diagonal crosses) gas data for NGC~5218.}\label{vel_comb.fig}
\end{figure}

\begin{figure}
\centerline{\includegraphics[clip=,angle=270,width=\columnwidth]{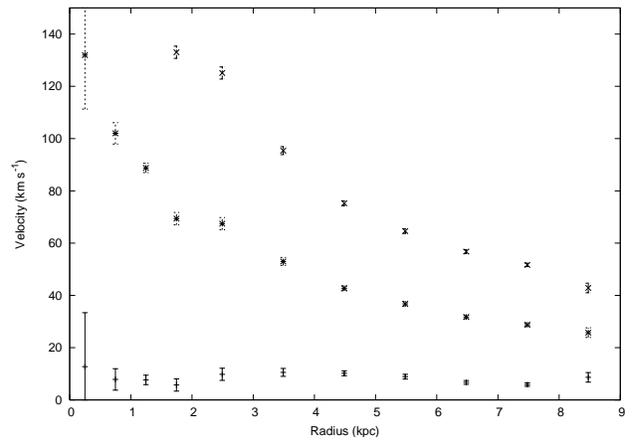}}
\caption{The angular velocity curve $\Omega(R)$ (stars) and derived curves
$\Omega + \kappa/2$ (diagonal crosses) and $\Omega - \kappa/2$ (plus signs),
using the model circular velocity from the combined molecular and atomic gas
datasets.}\label{omega_all.fig}
\end{figure}

The {\HI} rotation curve was fitted by breaking the galaxy into annuli $5''$ in
width (approximately the half-beam width) and fitting the velocity field in
each annulus, keeping the dynamical centre and inclination angle fixed. Radii
below $10''$ in which the {\HI} is seen predominately in absorption were
excluded. A similar fit was performed on the molecular gas data, using the same
fixed dynamical centre and inclination angle. For these higher resolution data
$2\farcs5$ annuli were used, and the data were fit from zero radius outwards.
In Fig.~\ref{vel_comb.fig} we plot the circular velocity $v(R)$, resulting from
analysis. We fit a function to these data the differential of which was
evaluated to allow computation of $\kappa$, the epicyclic frequency. In
Fig.~\ref{omega_all.fig} we plot the angular velocity, $\Omega(R) = v(R)/R$ and
$\Omega\pm\kappa/2$ curves. The error bars on Figs~\ref{vel_comb.fig} and
\ref{omega_all.fig} do not include any possible bias due to beam smearing.

\subsection{The length of the bar and co-rotation}\label{barlength.sec}

\begin{figure}
\centerline{\includegraphics[clip=,angle=270,width=\columnwidth]{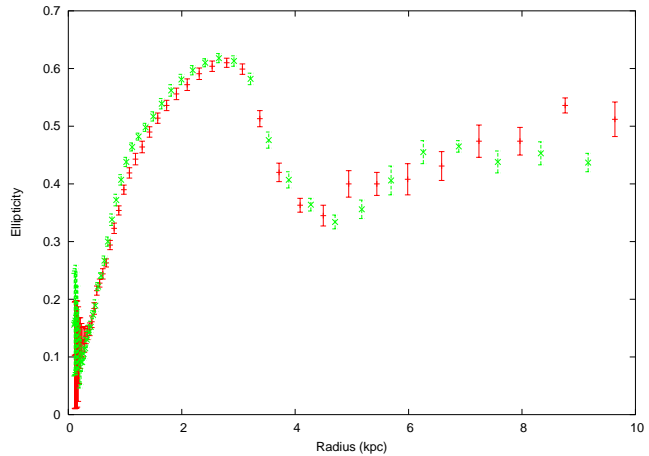}}
\caption{The ellipticity of NGC~5218 as a function of radius obtained from the
ellipse fitting of the IRAF STSDAS task ELLIPSE, for both the 2MASS $J$-band
(green) and $K$-band (red) images.}\label{ellip_fit.fig}
\end{figure}

Science-grade images in $J$, $H$ and $K_{\rm s}$ were downloaded from the Two
Micron All Sky Survey (2MASS) archive\footnote{Available at\\{\tt
http://irsa.ipac.caltech.edu/applications/2MASS/LGA/}}. The $J$, $H$ and
$K_{\rm s}$ passbands are less sensitive to extinction by dust in the central
regions of the galaxies than shorter wavelengths, and offer a bar tracer that
is unlikely to be contaminated by recent star formation. We adopt the method of
\cite{1997AJ....114..965R} -- see also \cite{2001ApJ...562..139M,
2002AJ....124.2581S} and \cite{2003ApJ...592L..13S} for examples of this
technique -- to estimate the length of the bar; this assumes the bar ends
correspond to the point of maximum ellipticity when fitting the galaxy
isophotes. The ellipticity of the 2MASS image isophotes, in the plane of the
sky, were fit using the IRAF STSDAS task ELLIPSE, which measures the surface
brightness, ellipticity and position angle of the semi-major axis.
Figure~\ref{ellip_fit.fig} shows the ELLIPSE fitted ellipticities for both the
$J$- and $K$-band 2MASS images. Both images give very similar results; the
$J$-band image indicates a maximum ellipticity of $0.618 \pm 0.008$ at a radius
of $13\farcs3$ and the $K$-band image a maximum ellipticity of $0.610 \pm
0.008$ at a radius of $14\farcs0$.

The bar's de-projected length $l$ can then be obtained from the observed length
$l_{\rm obs}$ using $l_{\rm obs} = l \sqrt{\cos^{2}\Phi +
\sin^{2}\Phi\cos^{2}i}$ \citep{1995AJ....110..199A}, where $i$ is the
inclination angle of the galaxy ($49\fdg5$) and $\Phi$ is the angle the bar
makes with galaxy's major axis in the plane of the galaxy. To determine $\Phi$
we take the position angle of the galaxy to be the average of position angles
fitted to the {\HI} data using the tilted-ring model in GAL and the position
angle of the bar to be that obtained by the IRAF package ELLIPSE at maximum
ellipticity. These angles are $71\fdg4 \pm 4\fdg3$ and $81\fdg6 \pm 0\fdg3$
respectively. Thus, the angle the bar makes with the major axis in the plane of
the sky $\Phi' = 10^\circ \pm 5^\circ$, and $\Phi$ follows since $\tan(\Phi') =
\tan(\Phi)\cos(i)$, thus $\Phi = 15^\circ \pm 5^\circ$. These angles yield a
very small correction factor in converting between the observed and
de-projected bar length: $l = 1.02 \times l_{\rm obs}$, giving $l=28\farcs6$
and $r = 14\farcs3$. Assuming co-rotation is at $1.2 \times R$
\citep{1992MNRAS.259..345A, 1994mtia.conf..143A} yields a co-rotation radius of
$17\farcs1$ which corresponds to a physical radius of 3.4~kpc and gives and
estimate of the pattern speed of the bar of 40 $\kms$ kpc$^{-1}$, which is
similar to that seen in other SB galaxies \citep[e.g.,][]{1995MNRAS.274..933M,
1999MNRAS.306..926G, 1998MNRAS.297.1052L} and in our Galaxy
\citep{1999ApJ...524L..35D}. Examining the angular velocity obtained from the
rotation curve fitting at a radius of 3.4~kpc we find the velocity is
significantly larger than the peak value of the $\Omega - \kappa/2$ curve; this
implies there is no inner Lindblad resonance in NGC~5218. We note that the
calculated radius of co-rotation (3.4~kpc) is not dissimilar to the radius
obtained by measuring distance between the molecular gas peaks toward the ends
of the bar (3.9~kpc).

Constraints on the existence or not of an inner Lindblad resonance in NGC~5218
depend sensitively on $\Omega(R)$ and $\kappa(R)$. As such they are subject to
a large degree of uncertainty. In particular, the absence of an observed peak
in the $\Omega-\kappa/2$ is symptomatic of a mass distribution which is not
strongly centrally condensed; inner Lindblad resonances only develop if the
mass distribution is sufficiently centrally peaked. It is well known that low
resolution observations smooth out the steep central rise of rotation curves
which results from the core and bulge of the galaxy
\citep{1999ApJ...523..136S}. It is possible that the limited resolution of our
observations is making it difficult to detect the steeply rising rotation curve
in the central region of NGC~5218 and as such yielding an artificially smooth
$\Omega-\kappa/2$ curve.

\subsection{Comparison of the {\HI} and CO dynamics}\label{5218_hicocomp.sec}

Figure~\ref{vel_comb.fig} shows the fitted circular velocities for both the
{\HI} and CO datasets. Around radii of approximately 3~kpc, where we have both
atomic and molecular gas observations there is a clear deviation in the
velocity structure of the two components.  This is not unexpected.  The fall in
the angular velocity of the molecular gas around 3~kpc, in the region of the
bar end may be a result of streaming motions within the bar. Gas on the
elongated, so-called $x_{1}$ orbits in the bar \citep{1994mtia.conf..143A}, has
a rotational velocity at the apex of the ellipse, near the end of the bar,
which is lower than that of gas on a circular orbits at the same radius,
therefore its $\Omega(R)$ will be lower. By comparison, if the {\HI} has a
large scale height than the molecular gas, it is likely to have been less
influenced by the bar in this region. We have also performed a more detailed
comparison of the {\HI} and CO dynamics by comparing the {\HI} and \CO{1}{0}
spectra at four locations across the disk. The spectra were extracted from data
cubes at a common resolution of $17\farcs2 \times 13\farcs5$. In the central
region the linewidths and lineshape of the spectra were very similar. In the
more extended regions the molecular gas displays a broader linewidth than the
{\HI}. In particular, at the eastern end of the bar, the molecular gas displays
a two-peaked profile with the atomic gas indicating emission in only one of
these peaks.

\section{Discussion and Conclusions}\label{discuss}

\subsection{Enhanced central molecular gas}\label{enhancecirc.sec}

In Section~\ref{gas_content.sec} we showed that NGC~5218 has a large molecular
gas mass for an SBb galaxy. The molecular gas distribution is strongly
centrally concentrated with a peak surface density greater than 2000
M$_{\odot}$ pc$^{-2}$ (Section~\ref{5218_co.sec}). Numerous numerical
simulations predict that bars, whether spontaneously or externally triggered,
can induce large scale gas inflow toward the central regions of galaxies
\citep[e.g., MH96;][]{1988A&A...203..259N,1991ApJ...370L..65B}.  Two recent
observational studies have found strong evidence for this process; both
\cite{1999ApJ...525..691S} and \cite {2005ApJ...632..217S} have demonstrated
that barred galaxies have a larger concentration of molecular gas in their
central kiloparsec than unbarred systems. \cite{1999ApJ...525..691S} used the
NRO--OVRO survey of 20 nearby spiral galaxies
\citep[see][]{1999ApJS..124..403S} to show that the degree of concentration of
molecular gas in the central kiloparsec of barred spiral galaxies is greater
than in their unbarred counterparts. They quantify this effect using the ratio
of the molecular gas surface density in the central kiloparsec to that over the
entire optical disk ($f_{\rm con}$), finding this ratio is on average four
times larger in barred than in unbarred systems. \cite{2005ApJ...632..217S}
studied a sample of 50 nearby spirals -- including both late type barred and
unbarred spirals, and CO-bright and CO-faint galaxies -- and found the effect
of the bar equally pronounced in late and early Hubble type spirals. Atomic gas
observations also indicate higher gas column densities in barred galaxies.
\cite*{1999ApJ...522..157R} examined the atomic gas column density in front of
Seyfert nuclei using X-ray measurements and found significantly higher column
densities in Seyferts with barred hosts than those in the unbarred sample. To
estimate $f_{\rm con}$ for NGC~5218 we remade the channel maps at the highest
possible spatial resolution ($\theta_{\rm FWHM} \sim 5''$, $\sim 1$ kpc).
Although beam smearing introduces uncertainty into our analysis, we find a
\CO{1}{0} flux over the central kiloparsec in NGC~5218 which corresponds to a
gas mass of $\sim 1.5 \times 10^{9}$ M$_{\odot}$. This is equivalent to a
molecular hydrogen surface density of $\sim 1300$ M$_{\odot}$ pc$^{-2}$ and a
gas surface density of $\sim 1800$ M$_{\odot}$ pc$^{-2}$. This is approximately
six times the mean of the barred sample of \cite{2005ApJ...632..217S} ($\sim
300$ M$_{\odot}$ pc$^{-2}$). In addition, the data for NGC~5218 indicate an
$f_{\rm con}$ value of 113, almost twice the average of the BIMA sample ($\sim
60$) and slightly larger than the average of the NRO--OVRO sample ($\sim 100$).
The discrepancy between the two literature values can be attributed to the
NRO--OVRO sample consisting mainly of CO-luminous galaxies. In contrast, the
BIMA sample was selected on the basis of $B$-band magnitude and thus
represents, at least from the perspective of CO flux comparison, a less biased
sample. The existence of the bar in NGC~5218 has led to a very large central
molecular gas surface density, even within the context of barred galaxies. For
comparison we note that typical mean densities of the largest molecular cloud
complexes in M31, M33 and M51 are in the range $40{-}500$ M$_{\odot}$
pc$^{-2}$, whilst the gas surface density of a typical giant molecular cloud in
the outer disk of the Milky Way is 200 M$_{\odot}$ pc$^{-2}$
\citep{1987ip...symp...21S}.

The impact of the bar potential on the gas distribution in NGC~5218 is
consistent with the bar's ellipticity, which implies it is a strong bar. The
ellipticity of a bar is thought to provide a good measure of the bar strength.
\cite{1995AJ....109.2428M} identify strong bars as those with an ellipticity
greater than 0.4; the 2MASS $K$-band data for NGC~5218 give a deprojected
ellipticity for NGC~5218 of the bar of at least 0.4. Simulations by
\cite{2004ApJ...600..595R} find that weak bars have almost no effect on the
radial distribution of the ISM. In contrast, strong extended bars are likely to
be more effective at driving gas toward the centre of the galaxy. Their
simulations also indicate that bars are only able to drive gas inwards to a
radius at which a ring forms; for bars containing inner rings, little net
inflow takes place inside this radius. Nuclear rings are likely to form when
so-called $x_{2}$ orbits exist; such orbits require the presence of two inner
Lindblad resonances. Our very high central gas densities are consistent with no
inner ring and no inner Lindblad resonances as our analysis of the bar dynamics
suggests.

\subsection{CO-to-H$_{2}$ conversion factor}

The molecular gas mass in NGC~5218 and the resulting gas column densities are
calculated using the CO-to-H$_{2}$ conversion factor, $X \equiv N({\rm
H_2})/I_{\rm CO} = 2.8 \times 10^{20}$ cm$^{-2}$ K~km~s$^{-1}$
\citep{1986A&A...154...25B}, which is appropriate for molecular gas clouds in
the Milky Way \citep{1984ApJ...276..182S}. In adopting this conversion factor
we assume that the conditions of the molecular gas clouds in NGC~5218 are
similar to the molecular gas clouds in the Milky Way. This is an assumption,
and it is possible that the CO emissivity in this system is enhanced due to
high pressures and/or velocity dispersion in the bulge region
\citep{2001ApJ...561..218R}. However, the observation of a $^{13}$CO
($J=1{-}0$)/\CO{1}{0} line ratio of $9.0 \pm 1.3$ \citep{1991A&A...249..323A}
in NGC~5218, which is comparable to the value observed in the Milky Way when
averaged over 300 pc, lends weight to the adoption of the Milky Way conversion
factor.

Although the molecular gas distribution in NGC~5218 is centrally concentrated,
there is also evidence of more extended emission (see
Fig.~\ref{5218co10_mom0.fig}). In particular, the \CO{1}{0} integrated emission
indicates two molecular gas condensations slightly offset from the bar ends.
The radius of these gas peaks (3.9~kpc) is similar to the calculated radius of
co-rotation (3.4~kpc), see Section~\ref{barlength.sec}. The presence of these
gas complexes, like that of the large central peak, can be attributed to the
existence of a barred potential. Gas concentrations at bar ends have been
observed in a number of galaxies \citep[e.g.,][]{1991ApJ...381..118K,
1996ApJ...461..186D, 2000ApJ...532..221S, 2002AJ....124.2581S}. These
concentrations may be stagnation points in the gas flow; gas on $x_{1}$ orbits
at the bar ends will have a smaller angular velocity than more central gas due
to its larger radius. Their presence may also be due to the convergence of gas
on elliptical orbits in the bar with gas on more circular orbits in the region
just beyond the bar \citep{1991ApJ...381..118K}; these gas complexes are
located in a transition region between the spiral arms and bar ends. Although
the bar in NGC~5218 appears to have contributed to a central concentration of
gas, this process appears to be ongoing.

We now consider the origin of NGC~5218's high molecular gas mass. In
Section~\ref{gas_content.sec} we showed that NGC~5218 has a comparatively low
{\HI} mass. The bar may have led predominately to the transport of existing
molecular material from the disk to the centre of NGC~5218, or this transport
could have involved both atomic and molecular gas with subsequent conversion of
atomic to molecular gas in the central region. \cite{2004A&A...416..515D}
studied the effects of bars on the neutral hydrogen content of starburst and
Seyfert galaxies and found that the barred galaxies had a lower total {\HI}
mass than unbarred galaxies. They suggest that atomic gas is funnelled toward
the centre of barred galaxies where it is then converted to molecular gas. In
Section~\ref{a104_hi.sec} we used {\HI} absorption observations to estimate the
atomic gas column density in the centre of NGC~5218 to be $(3.0 \pm 0.3) \times
10^{21}$ cm$^{-2}$. Various studies indicate a `transition' column density at
which {\HI} is sufficiently dense for self-shielding to allow the formation of
H$_{2}$ of $\sim 5\times 10^{20}$ cm$^{-2}$, increasing to $\geq 10^{21}$
cm$^{-2}$ for large molecular clouds \citep{1977ApJ...216..291S,
1979ApJ...227..466F, 1986PASP...98.1076F, 1994ApJ...429..672R}. The {\HI}
column density observed at the centre of NGC~5218 lies above these values so
{\HI} to H$_{2}$ conversion may be taking place at the centre of NGC~5218. The
high H$_{2}$ to {\HI} ratio in the centre of NGC~5218 and the dynamical
consistency of the two components in the central region (see
Section~\ref{5218_hicocomp.sec}) is certainly consistent with inflow of both
components and efficient {\HI} to H$_{2}$ conversion.

\subsection{Star formation}\label{starform.sec}

In Section~\ref{5218_starformrate.sec} we estimated the SFR of NGC~5218 using
both radio and FIR data; these data yield SFRs of 3.4 and 5.3 M$_{\odot}$
yr$^{-1}$ respectively. A large proportion of the star formation is taking
place in the dusty central region of NGC~5218 (see below) where we observe a
very large molecular gas density. However, both the radio and FIR provide
largely extinction-free probes of the star formation and as such should yield
robust estimates of the SFR. Comparison of the NVSS ($\theta_{\rm FWHM}=45''$)
flux density ($30.4\pm1.0$ mJy) and the FIRST \citep*{1994ASPC...61..165B}
($\theta_{\rm FWHM}=5\farcs4$) flux density ($23.06\pm0.14$ mJy) suggests at
least 75\% of the galaxy's star formation is coming from the central kiloparsec
region.

Before discussing the star-formation properties of NGC~5218 we note a potential
caveat with the estimation of the SFR of this system. A number of authors have
examined the optical spectroscopy of NGC~5218 for AGN activity. On the basis of
their investigation \cite*{1995ApJ...447..545A} conclude the emission line
properties of NGC~5218 are ambiguous, and \cite{1995ApJS...98..171V} classify
it as a LINER. NGC~5218 sits comfortably on the radio-FIR correlation
\citep*{2001ApJ...554..803Y} with no obvious radio or FIR excess; the two SFR
estimators yield very similar results (Section~\ref{5218_starformrate.sec}). In
addition, recent work by \cite{2005ApJ...620..113D} suggests that in their
sample of IR-bright LINERS the low X-ray luminosities and Eddington ratios
imply the contribution from the AGN to the FIR luminosity is likely to be
small. However, we cannot exclude the possibility that nuclear activity in this
system leads to an overestimated FIR SFR.

\subsubsection{Is the level of star formation large?}

NGC~5218 appears to have a substantial star formation rate. The 60-$\mu$m
luminosity of NGC~5218, $1.5\times10^{10}$ L$_{\odot}$, is more than three
times the median for Sb--Sbc galaxies found by \cite{1997AJ....113..599D},
although it does not qualify it as Luminous Infrared Galaxy (LIRG). NGC~5218's
$H$-band normalised SFR, an approximation of star formation per unit mass,
implies a slightly enhanced value \citep{1997AJ....113..599D}. The mean value
of $\log(L_{\rm FIR}/L_{\rm B})$ obtained by \cite{1997AJ....113..599D} for
Sb--Sbc galaxies was $-0.48\pm0.34$, whereas the value for NGC~5218 is $-0.05$,
consistent with a high current SFR in relation to its recent star formation
history (although it may also reflect, to some extent, increased gas and
therefore dust in its centre).

The star-formation efficiency (SFE) is the ratio of the current star-formation
rate to the mass of molecular gas: for NGC~5218 the SFE derived from the FIR
star formation rate and total molecular gas mass is $\sim 8\times10^{-10}$
yr$^{-1}$. This value, whilst larger than the average for Sb galaxies
\citep{2002A&A...385..412M}, is still significantly below the $\sim 30\%$ over
$10^8$~yr found in starburst systems \citep{1998ARA&A..36..189K}. Despite
NGC~5218's relatively high star-formation rate, its global star-formation
efficiency appears to be more in keeping with that of a normal Sb galaxy than a
starburst system. The star formation efficiency of the central $2{-}3$ kpc is
approximately three times the global value, falling toward the lower end of
those systems classified as nuclear starbursts \cite*{1997A&A...325...81P}

\cite{1999ApJ...525..691S} have investigated the gas-to-dynamical mass ratio in
the central kiloparsec of a sample of 20 nearby spiral galaxies. They observe a
larger ratio in galaxies with an {\HII} spectral classification, and interpret
this as evidence for more active star formation and nuclear spectra dominated
by {\HII} regions in galaxies which have large amounts of gas given their size.
They suggest that the effect may be related to the stability of the nuclear
molecular gas disk, relating $M_{\rm gas}/M_{\rm dyn}$ to the Toomre $Q$
parameter \citep{1964ApJ...139.1217T}. For comparison with these data we use
the highest resolution CO data set for NGC~5218 to examine the gas-to-dynamical
mass ratio in central kiloparsec. We integrate the flux obtained in the central
molecular gas peak and compare the mass obtained with an estimate of the
dynamical mass in this region. We select a physical region with a radius of $r
\sim 1$ kpc, larger than the region ($r = 500$ pc) examined by
\cite{1999ApJ...525..691S} because we can make a more robust estimate of
$M_{\rm gas}/M_{\rm dyn}$ on this scale given our available data. The region
examined lies well within the physical extent of the bar; the molecular gas
dynamics in this region are dominated by circular motion
(Fig.~\ref{5218co10_mom1.fig}). Assuming Keplerian rotation and a spherical
mass distribution, the dynamical mass is
\[
  \left(\frac{M_{\rm dyn}}{\rm M_{\odot}}\right) =
   2.3\times10^{5} \left(\frac{r}{{\rm kpc}}\right)
                   \left(\frac{v}{\kms}\right)^{2},
\]
where $v$ is the rotational velocity in the plane of the disk at a radius $r$.
The 20\% linewidth over the 1.1~kpc radius region is 280 $\kms$. Correcting for
the inclination of the galaxy yields a rotational velocity in the plane of the
disk of 184 $\kms$, corresponding to a dynamical mass of $8.5\times10^{9}$
M$_{\odot}$. Integrating the CO flux observed over this area yields a molecular
hydrogen mass of $1.6\times10^{9}$ M$_{\odot}$ and a molecular gas mass of
$2.2\times10^{9}$ M$_{\odot}$. The molecular gas mass fraction over the central
1.1~kpc radius region in NGC~5218 is 26\%, $1\sigma$ above the value obtained
by \cite{1999ApJ...525..691S} for {\HII} dominated systems ($17.6 \pm 8.0\%$
over the central kiloparsec). If we reduce the physical scale over which we
undertake the comparison, the mass fraction increases dramatically. Our data
imply that the gas-to-dynamical mass ratio over the central region of NGC~5218
is large when compared with actively star forming {\HII} systems in the
Sakamoto et al.\ sample.

We note that the derived $M_{\rm gas}/M_{\rm dyn}$ is only related to the
Toomre $Q$ parameter and the central gas stability if the gas is in the form of
a nuclear disk. Bar-streaming motions introduce inherent uncertainty into the
use of the ratio $M_{\rm gas}/M_{\rm dyn}$ as a probe of the stability. In
addition, many theorists are sceptical about using the Toomre $Q$ parameter as
a measure of instability toward star formation, arguing that feedback from such
star formation would lower the parameter on short timescales (e.g.,
\citealt{2001ASPC..249..475C}, see also \citealt{2005A&A...431..887K}).
Nonetheless, the results of \cite{1999ApJ...525..691S} indicate that all except
one of the {\HII} galaxies in their sample have an $M_{\rm gas}/M_{\rm dyn}$
greater than 0.1 and we believe comparison with these observations is useful.

\subsection{Bar formation in NGC~5218}

Both simulation-based and observational evidence exists linking tidal
perturbations with the triggering of bars. In particular, observations in the
literature point to bars in early-type galaxies, of which NGC~5218 is an
example, being produced in this way. This result is supported by simulations
which indicate that a bulge component acts to stabilise the galaxy against
spontaneous bar formation, an increasingly massive bulge offering increased
stability (MH96). It may be that for galaxies with bulges an external trigger
is a prerequisite for a bar to form. To investigate any link between the bar in
NGC~5218 and its tidal interaction, we look in more detail at the nature of the
bar, undertaking comparisons with both simulation-based and observational data.

Simulations suggest that interactions can affect bar formation in galaxies in a
number of different ways: they can trigger bar formation in an otherwise stable
galaxy; speed up the process of bar formation in a system which is already
unstable \citep{1990A&A...230...37G}; regenerate a dissolved or weakened bar
\citep{2004MNRAS.347..220B} and enhance or decrease the strength of an existing
bar \citep{1990A&A...230...37G}. All such scenarios are possible in NGC~5218,
although the strength of the bar, as measured from the axial ratio, would imply
that the tidal encounter is unlikely to have significantly weakened the bar.

In an investigation of the blue and near-infrared surface photometry of 15
barred spirals, \cite{1985ApJ...288..438E} observed two different types of
major axis surface density profiles, exponential and flat. On the basis of
their simulations, \cite{1996ApJ...469..605N} argue that exponential profiles
are found in spontaneously produced bars, whilst flatter profiles occur in bars
produced by tidal instability. Unfortunately the photometric data available for
NGC~5218 are not of sufficiently high resolution to confidently discern between
the two profiles; the large central peak observed in all bars yields a profile
that appears exponential when smoothed to lower resolution. Nonetheless, the
SDSS $z$-band light profile indicates some flattening and a flattened profile
is certainly possible. In Noguchi's interpretation, this implies the bar
profile is consistent with a bar produced by tidal perturbation. Recent work by
\cite{2004MNRAS.347..220B} has questioned this interpretation, arguing that the
results of \cite{2002MNRAS.330...35A} indicate that both profiles occur in
isolated systems and that the different profiles are associated with varying
bar strengths rather than the triggering mechanism, flatter profiles being
produced in stronger bars.

Our analysis of NGC~5218 gave an estimated pattern speed for the bar of $\sim
40$ $\kms$ kpc$^{-1}$. Studies by \cite{1998ApJ...499..149M} show that in weak
tidal perturbations the properties of the bar depend largely on the internal
structure of the host galaxy, rather than the speed of the perturber (see also
simulations by \citealt{1990A&A...230...37G}). However, for stronger tidal
perturbations the pattern speed is dependent on the perturber mass (see also
\citealt{2004MNRAS.347..220B}). \citeauthor{1998ApJ...499..149M} argue that for
the strongest perturbations tidally induced bars can rotate slowly with an
inner Lindblad resonance near the bar ends, whereas bars produced by weaker
perturbations have no inner-Lindblad resonances, rotate more rapidly, and end
near co-rotation. In the case of NGC~5218 we inferred that there are no
inner-Lindblad resonances, and that if the bar-ends are co-rotating its angular
velocity is $\sim 40$ $\kms$ kpc$^{-1}$. However, since the $\Omega - \kappa/2$
curve provides a relatively constant resonance throughout the galactic radii at
a velocity of 10 $\kms$ kpc$^{-1}$, we cannot exclude the possibility that the
bar in NGC~5218 is rotating at this, much slower velocity, and ending before
co-rotation.

A number of investigations of tidal triggering of bar formation have looked at
the relationship between strength of interaction and bar formation
\citep{1998ApJ...499..149M,2004MNRAS.347..220B}. Quantitative comparison
between these simulations and our results is difficult since many of the
results are sensitive to the mass concentration of the disk and the
disk-to-halo mass fraction which are difficult to constrain observationally.
Nonetheless, we can make some important observations about this system. Given
the apparently low mass ratio of NGC~5216 to NGC~5218 ($1:2$ using $K$-band
data, Section~\ref{104_sim.sec}), the perigalactic separation required to
produce a tidal perturbation of sufficient strength to induce bar formation is
likely to be small. The necessity of a small perigalactic separation is
consistent with the simulations, which indicate that production of the extended
tidal features observed in the {\HI} distribution require a small perigalactic
separation.

The gas surface density in the central kiloparsec of NGC~5218 (1300 M$_{\odot}$
pc$^{-2}$) is significantly larger than the mean of both the unbarred ($107 \pm
29$ M$_{\odot}$ pc$^{-2}$) and barred sample ($309 \pm 71$ M$_{\odot}$
pc$^{-2}$) of \cite{2001ASPC..249..605S}. The enhanced surface density observed
in NGC~5218 can be accounted for by an additional molecular gas mass of the
order of $9\times10^{8}$ M$_{\odot}$. Estimates of gas inflow rates in barred
galaxies vary widely \citep[$0.1{-}17$ M$_{\odot}$
yr$^{-1}$,][]{1993A&A...268...65F, 1995ApJ...441..549Q, 1998MNRAS.297.1052L,
1997ApJ...482L.143R, 1999ApJ...525..691S, 2004ApJ...616..199I}; adopting the
value inferred by \cite{1995ApJ...441..549Q} of $4 \pm 2$ M$_{\odot}$ yr$^{-1}$
via calculation of the torques on the molecular gas using near-IR images, we
find the time taken to produce the observed enhanced central gas density in
NGC~5218 is similar to the time since perigalacticon. However, it must be
remembered the inflow rate is not certain, and may vary with time.

\subsection{Conclusions}

NGC~5218 is a barred interacting spiral galaxy which has undergone a prograde
encounter with an elliptical, NGC~5216, sometime in the recent past. Both the
{\HI} velocity field and the simulations indicate the timescale since
perigalacticon is of the order of $3 \times 10^{8}$ yr. NGC~5218 has a very
large central molecular gas surface density when compared to the molecular gas
surface density in the central kiloparsec of the BIMA sample of unbarred
galaxies \citep{2001ASPC..249..605S}. The difference can be accounted for if we
assume gas inflow to the central region over the time since perigalacticon and
at a rate comparable to those inferred in other systems
\citep[e.g.,][]{1995ApJ...441..549Q}. Our observations are consistent with the
interaction of NGC~5218 with NGC~5216 triggering an instability in the galactic
potential or enhancing an existing instability that has led to gas inflow
yielding a central molecular gas density at the elevated level observed.

The nature of the bar, molecular gas and dynamical time scales we infer for NGC
5218 are very similar to those in UGC 2855, as discussed by
\cite{1999A&A...346...45H}.  This interacting galaxy also has a high central
gas concentration with no evidence for any inner Lindblad resonance or shocking
of the molecular gas.  H\"{u}ttemeister et al.\ deduce that the bar is very
young and the system is in a pre-starburst phase.

NGC~5218 has a relatively large SFR (Section~\ref{starform.sec}) It has a
normal disk-wide star-formation efficiency although the nuclear region has a
star-formation efficiency toward the lower end of starburst nuclei. It is
possible that NGC~5218 is in the early stages of bar-driven evolution and has
not yet reached its period of most active star formation. Whilst the bar in
NGC~5218 appears to have contributed to large central molecular gas density
there is still a significant fraction of the molecular material found in the
more extended regions of this system. The process of gas transfer toward the
central region of NGC~5218 is clearly ongoing. The molecular gas density in the
centre of NGC~5218 may not yet be sufficiently large to have triggered a
starburst phase. We suggest that NGC~5218 is likely to undergo a period of more
enhanced star formation in the near future. This proposition is supported by
the following evidence.
\begin{itemize}
\item The large concentration of molecular gas and very high surface density in
the central region of NGC~5218.
\item A sizable, extended molecular gas component which indicates that the
process of gas transfer to the central region is ongoing.
\item NGC~5218's large FIR-luminosity and SFR when compared with other galaxies
of the same Hubble type.
\item The large $L_{\rm IR}/L_{\rm B}$ ratio, indicating that the
current-to-recent star formation in NGC~5218 is larger than an average
early-type spiral.
\item The large $M_{\rm gas}/M_{\rm dyn}$ ratio in the centre of NGC~5218
implying a large quantity of gas in this region given its size. This may be
indicative of instability toward star formation.
\end{itemize}

\section*{Acknowledgements}

The James Clerk Maxwell Telescope is operated by The Joint Astronomy Centre on
behalf of the Particle Physics and Astronomy Research Council of the United
Kingdom, the Netherlands Organisation for Scientific Research, and the National
Research Council of Canada. We thank the staff of the GMRT that made these
observations possible. The GMRT is run by the National Centre for Radio
Astrophysics of the Tata Institute of Fundamental Research. The National Radio
Astronomy Observatory is a facility of the National Science Foundation operated
under cooperative agreement by Associated Universities, Inc. We thank the staff
of the OVRO interferometer. This research has made use of the NASA/IPAC
Extragalactic Database (NED) which is operated by the Jet Propulsion
Laboratory, California Institute of Technology, under contract with the
National Aeronautics and Space Administration. Funding for the SDSS and SDSS-II
has been provided by the Alfred P.\ Sloan Foundation, the Participating
Institutions, the National Science Foundation, the U.S. Department of Energy,
the National Aeronautics and Space Administration, the Japanese Monbukagakusho,
the Max Planck Society, and the Higher Education Funding Council for England.
The SDSS Web Site is {\tt http://www.sdss.org/}. The SDSS is managed by the
Astrophysical Research Consortium for the Participating Institutions. The
Participating Institutions are the American Museum of Natural History,
Astrophysical Institute Potsdam, University of Basel, Cambridge University,
Case Western Reserve University, University of Chicago, Drexel University,
Fermilab, the Institute for Advanced Study, the Japan Participation Group,
Johns Hopkins University, the Joint Institute for Nuclear Astrophysics, the
Kavli Institute for Particle Astrophysics and Cosmology, the Korean Scientist
Group, the Chinese Academy of Sciences (LAMOST), Los Alamos National
Laboratory, the Max-Planck-Institute for Astronomy (MPIA), the
Max-Planck-Institute for Astrophysics (MPA), New Mexico State University, Ohio
State University, University of Pittsburgh, University of Portsmouth, Princeton
University, the United States Naval Observatory, and the University of
Washington. The Digitized Sky Surveys were produced at the Space Telescope
Science Institute under U.S.\ Government grant NAG W-2166. The images of these
surveys are based on photographic data obtained using the Oschin Schmidt
Telescope on Palomar Mountain and the UK Schmidt Telescope. The plates were
processed into the present compressed digital form with the permission of these
institutions. The Second Palomar Observatory Sky Survey (POSS-II) was made by
the California Institute of Technology with funds from the National Science
Foundation, the National Geographic Society, the Sloan Foundation, the Samuel
Oschin Foundation, and the Eastman Kodak Corporation. This publication makes
use of data products from the Two Micron All Sky Survey, which is a joint
project of the University of Massachusetts and the Infrared Processing and
Analysis Center/California Institute of Technology, funded by the National
Aeronautics and Space Administration and the National Science Foundation. HC
acknowledges receipt of a PPARC studentship.


\label{lastpage}

\end{document}